\documentclass[twocolumn,showpacs,amssymb,prd]{revtex4}

\usepackage{amsmath}
\usepackage{latexsym}
\usepackage{graphicx}
\usepackage{epstopdf}

\begin{document}


\title{A new approach to the  Newman-Penrose formalism} 


\author{Andrea Nerozzi$\footnote{Electronic address: andrea.nerozzi@ist.utl.pt}
$}
\vskip 3mm
\affiliation{Centro Multidisciplinar de Astrof\'{\i}sica -
CENTRA, Dept. de F\'{\i}sica, Instituto
Superior T\'ecnico, Av. Rovisco Pais 1, 1049-001 Lisboa, Portugal}
%
%


\date{\today}


\begin{abstract}
The Newman-Penrose formalism in transverse tetrads, namely those tetrads where
$\Psi_1=\Psi_3=0$, is studied. In particular it is shown that the equations governing the
dynamics within this formalism can be recast in a particularly compact way, leading
to a better understanding of the formalism itself. The particular choice of tetrad allows 
not only to obtain the expression of Weyl scalars as simple functions of curvature invariants,
but also the spin coefficients can be partly determined in an invariant way, by means
of a new expression for the Bianchi identities that shows the various degrees of freedom
in a more intuitive and direct way. 
We expect this approach to be very promising for
a better understanding of all the equations governing the Newman-Penrose formalism.
Such a new insight to the equations can also turn out to be useful for a generalization of the Newman-
Penrose formalism to higher dimensions, thus allowing a better analysis of the 
various degrees of freedom, in view of extending to this case results already known in
four dimensions.
\end{abstract}




\maketitle

\section{Introduction}
\label{sec:intro}

The Newman-Penrose (NP) formalism is an invaluable tool in general relativity. It has been
used in several aspects of analytical and computational relativity: for example it is the only approach
that allows the determination of a single equation describing the perturbations of a rotating
black hole, through the Teukolsky equation \cite{Teukolsky73}. Furthermore it is
used for extracting gravitational waves from numerical simulation by calculating the
quantity $\Psi_4$. In order to obtain a well defined expression, i.e. consistent with
perturbation theory, for the scalar $\Psi_4$, it is necessary to choose the
null tetrad requiring that it converges to the
Kinnersley tetrad \cite{Kinnersley69} when the space-time approaches Petrov type D, i.e. the single black 
hole space-time, which is the end state of the scenarios normally considered
in numerical relativity. This tetrad has been dubbed {\it quasi-Kinnersley
tetrad}.
Refs \cite{beetle-2005-72,nerozzi-2005-72} show that the {\it quasi-Kinnersley tetrad} belongs
to a group of tetrads, the {\it quasi Kinnersley frame}, whose elements are connected
to each other by spin/boost (type III) transformations. One possible
{\it quasi-Kinnersley frame} was found to be
one of the three transverse frames where $\Psi_1=\Psi_3=0$.

The {\it quasi-Kinnersley frame} \cite{beetle-2005-72, nerozzi-2005-72, nerozzi-2006-73, Nerozzi:2006aj, campanelli-2006-73b} has been shown to be of great importance for 
wave extraction. In fact, when computed in this particular frame, Weyl scalars are
directly associated with the relevant physical degrees of freedom, thus giving 
a better characterization of the physical properties of the scenario being studied. 
In Ref. \cite{nerozzi-2007-75} we have shown that it is possible to give an explicit expression for the
Weyl scalars in the {\it quasi-Kinnersley} frame once a preferred time-like observer
is identified. However, these expressions are valid in the case where the two scalars
$\Psi_0$ and $\Psi_4$ coincide, corresponding to a specific choice of the spin/boost 
parameter identifying the type III rotation. Unfortunately, this choice is not good for wave
extraction, as the coinciding Weyl scalars would show in this case an incorrect peeling
fall-off at large distances from the source. In a follow up paper \cite{Nerozzi:2008ng} we have shown
that by means of the Ricci and Bianchi identities, it is possible to improve this result 
and calculate the optimal spin/boost parameter that gives $\Psi_0$ and $\Psi_4$ with
the correct peeling fall-off, by imposing the condition that the spin coefficient
$\epsilon$ goes to zero in the limit of the Kinnersley tetrad. This calculation however
has been made assuming a Petrov type D space-time,  
therefore the final expression is dependent
on background parameters of the single black hole and not expressed in terms of 
generally defined curvature invariants. 
In view of extending this result to a general Petrov type I space-time, we show in this paper that this study can be carried further, characterizing
all the fundamental quantities that are introduced in the Newman-Penrose formalism,
and giving a precise meaning to some of the equations introduced in this
formalism, namely the Bianchi identities, that become eliminant conditions to fix some 
of the relevant degrees of freedom in the formalism. This is
achieved by defining new vector quantities, from which spin coefficients can be
calculated easily, that are invariant under specific tetrad transformations and hence
the possibility to express them in function of curvature invariants.
The aim of this work is thus to present the Newman-Penrose formalism in a new way 
introducing only gauge invariant quantities and reformulating the relevant equations only
in function of those; this paper can be thought as half way through, giving the
expression of the Bianchi identities within this new approach. In section \ref{sec:weyl} we will introduce the general notation adopted by this paper while in section \ref{sec:newvect}
we will define this new set of gauge invariant vectors in function of which the Bianchi identities can be reformulated in
a new compact form. 

\section{Weyl scalars, spin coefficients and Bianchi identities}
\label{sec:weyl}

The relevant quantities in the NP formalism are the Weyl scalars, defined
as

\begin{subequations}
\label{eqn:weylscalars}
\begin{eqnarray}
\Psi_0 &=&-C_{abcd}\ell^am^b\ell^cm^d, \label{eqn:weylscalars0} \\
\Psi_1 &=&-C_{abcd}\ell^an^b\ell^cm^d, \label{eqn:weylscalars1} \\
\Psi_2 &=&-C_{abcd}\ell^am^b\bar{m}^cn^d, \label{eqn:weylscalars2} \\
\Psi_3 &=& -C_{abcd}\ell^an^b\bar{m}^cn^d, \label{eqn:weylscalars3} \\
\Psi_4 &=& -C_{abcd}n^a\bar{m}^bn^c\bar{m}^d, \label{eqn:weylscalars4} 
\end{eqnarray}
\end{subequations}
and the connection coefficients (spin coefficients), given by

\begin{subequations}
\label{eqn:spicoeffdef}
\begin{eqnarray}
\rho&=&m^{\mu}\bar{m}^{\nu}\nabla_{\nu}\ell_{\mu},  \\
\lambda&=&n^{\mu}\bar{m}^{\nu}\nabla_{\nu}\bar{m}_{\mu}, \\
\epsilon&=&2^{-1}\cdot \ell^{\nu}\left(n^{\mu}\nabla_{\nu}\ell_{\mu}+m^{\mu}\nabla_{\nu}\bar{m}_{\mu}\right),\\ 
\mu&=&n^{\mu}m^{\nu}\nabla_{\nu}\bar{m}_{\mu},\\
\sigma&=&m^{\mu}m^{\nu}\nabla_{\nu}\ell_{\mu},  \\
\gamma&=&2^{-1}\cdot n^{\nu}\left(n^{\mu}\nabla_{\nu}\ell_{\mu}+m^{\mu}\nabla_{\nu}\bar{m}_{\mu}\right),  \\ 
\tau&=&m^{\mu}n^{\nu}\nabla_{\nu}\ell_{\mu}, \\
\nu&=&n^{\mu}n^{\nu}\nabla_{\nu}\bar{m}_{\mu},  \\
\beta&=&2^{-1}\cdot m^{\nu}\left(n^{\mu}\nabla_{\nu}\ell_{\mu}+m^{\mu}\nabla_{\nu}\bar{m}_{\mu}\right), \\
\pi&=&n^{\mu}\ell^{\nu}\nabla_{\nu}\bar{m}_{\mu},  \\
\kappa&=&m^{\mu}\ell^{\nu}\nabla_{\nu}\ell_{\mu},  \\
\alpha&=&2^{-1}\cdot \bar{m}^{\nu}\left(n^{\mu}\nabla_{\nu}\ell_{\mu}+m^{\mu}\nabla_{\nu}\bar{m}_{\mu}\right),  
\end{eqnarray}
\end{subequations}
where $C_{abcd}$ is the Weyl tensor and $\left[\ell^{\mu},n^{\mu}, m^{\mu}, \bar{m}^{\mu}\right]$
is the Newman-Penrose null tetrad.

The relevant equations 
are the Ricci and Bianchi identities written in terms of Weyl scalars and spin coefficients. 
The advantage of using the Newman-Penrose formalism relies upon the fact that the
gauge freedom in such a formalism corresponds to the choice of tetrad, as opposed to
general coordinate transformations. The gauge group, then, is the Lorentz group, on
which we have deeper theoretical knowledge. In Refs. \cite{beetle-2005-72,nerozzi-2005-72} it has been shown that given an algebraically general
space-time (Petrov type I) it is always possible to choose a tetrad where the two Weyl
scalars $\Psi_1$ and $\Psi_3$ vanish. This tetrad is not unique, and we have a clear knowledge
of the properties of such tetrads. In particular, we know that one of these tetrads converges naturally
to the Kinnersley tetrad in the limit of Petrov type D space-time (which is not surprising, as the Kinnersley
tetrad also has $\Psi_1=\Psi_3=0$). This can also be explained with the fact that the
null vectors constituting the transverse tetrads happen to be ``in the middle'' of
the two couples of principal null directions that eventually coincide in the limit of Petrov
type D (for a more rigorous explanation of this statement see \cite{nerozzi-2005-72}). It is well known that the principal null directions can give specific physical information about 
the space-time under study, hence one expects that a method that relates to 
principal null direction is better suited for extracting physical information. Recent works
\cite{Campanelli:2008dv, Owen:2010vw} have used the concept of principal null directions to extract relevant physical properties
from numerically evolved space-times.

The condition $\Psi_1=\Psi_3=0$ leaves an indetermination in the choice of the spin/boost
parameter. If we impose the additional condition that $\Psi_0=\Psi_4$ we also fix this last
parameter and the tetrad is completely fixed. The only remaining degrees of freedom are
$\Psi_2$ and $\Psi_4$, which can be written in this particular tetrad as $\Psi_2=-\frac{1}{2\sqrt{3}}\Psi_+$ and $\Psi_4=-\frac{i}{2}\Psi_-$, where

\begin{equation}
\Psi_{\pm}= I^{\frac{1}{2}}\left(e^{\frac{2\pi i k}{3}}\Theta\pm e^{-\frac{2\pi i k}{3}}\Theta^{-1}\right).
\label{eqn:pispm}
\end{equation}
In Eq.~(\ref{eqn:pispm}) $\Theta=\sqrt{3}P I^{-\frac{1}{2}}$, $P=\left[J+\sqrt{J^2-\left(I/3\right)^3}\right]^{\frac{1}{3}}$ and $I$ and $J$ are the two curvature invariants; $k$ is an integer number 
that spans the interval $\left[0,1,2\right]$ and identifies the three different transverse frames.

It is not surprising that in this tetrad $\Psi_2$ and
$\Psi_4$ are functions of curvature invariants only: having fixed all the gauge
degrees of freedom, we are left with the four degrees of freedom given by the curvature invariants,
hence any non-vanishing Weyl scalar must be a function of those.

For our study however we want to leave the spin/boost parameter not fixed. This is because,
as already mentioned, 
the choice $\Psi_0=\Psi_4$ is obviously wrong for the point of view of wave extraction, as the
two scalars would not have the correct fall-off dictated by the peeling theorem. This means that
$\Psi_0$ and $\Psi_4$ can be, in general, different. By defining the spin-boost parameter
as $\mathcal{B}=\left(\frac{\Psi_4}{\Psi_0}\right)^{\frac{1}{4}}$, the three non vanishing Weyl
scalars can be written as

\begin{subequations}
\label{eqn:psi2psi4qkt}
\begin{eqnarray}
\Psi_0&=&-\frac{i\mathcal{B}^{-2}}{2}\cdot \Psi_{-},
\label{eqn:psi0qkt} \\
\Psi_2&=&-\frac{1}{2\sqrt{3}}\cdot\Psi_{+},
\label{eqn:psi2qkt} \\
\Psi_4&=&-\frac{i\mathcal{B}^{2}}{2}\cdot \Psi_{-}.
\label{eqn:psi4qkt} 
\end{eqnarray}
\end{subequations}

The curvature invariants $I$ and $J$ can be expressed, in terms of the $\Psi_{\pm}$
scalars, as

\begin{subequations}
\label{eqn:curvinvariantsIJ}
\begin{eqnarray}
I&=&\frac{1}{4}\left(\Psi_+^2-\Psi_-^2\right), \\
J&=&-\frac{\Psi_+}{24\sqrt{3}}\left(\Psi_+^2+3\Psi_-^2\right).
\end{eqnarray}
\end{subequations}

In this paper we will study how such a choice of tetrad fixes the values of the other
relevant variables in the Newman-Penrose formalism, namely the spin coefficients. In order to do so, we write first the Bianchi identities in terms of the non-vanishing scalars,
thus obtaining

\begin{subequations}
\label{eqn:bianchimod}
\begin{eqnarray}
D\Psi_{+}&=&-\tilde{\lambda}\Psi_{-}+3\rho\Psi_{+},
\label{eqn:bianchimodD4} \\
D\Psi_{-}&=&\tilde{\lambda}\Psi_{+}-\left(4\tilde{\epsilon}-\rho\right)\Psi_{-},
\label{eqn:bianchimodD2} \\
\Delta\Psi_{+}&=&\tilde{\sigma}\Psi_{-}-3\mu\Psi_{+},
\label{eqn:bianchimodDelta4} \\
\Delta\Psi_{-}&=&-\tilde{\sigma}\Psi_{+}+\left(4\tilde{\gamma}-\mu\right)\Psi_{-},
\label{eqn:bianchimodDelta2} \\
\delta\Psi_{+}&=&-\tilde{\nu}\Psi_{-}+3\tau\Psi_{+},
\label{eqn:bianchimoddelta4} \\
\delta\Psi_{-}&=&\tilde{\nu}\Psi_{+}-\left(4\tilde{\beta}-\tau\right)\Psi_{-},
\label{eqn:bianchimoddelta2} \\
\delta^*\Psi_{+}&=&\tilde{\kappa}\Psi_{-}-3\pi\Psi_{+},
\label{eqn:bianchimoddeltas4} \\
\delta^*\Psi_{-}&=&-\tilde{\kappa}\Psi_{+}+\left(4\tilde{\alpha}-\pi\right)\Psi_{-},
\label{eqn:bianchimoddeltas2} 
\end{eqnarray}
\end{subequations}
where we have introduced the rescaled spin coefficients $\tilde{\lambda}=i\sqrt{3}\lambda\mathcal{B}^{-2}$,
$\tilde{\sigma}=i\sqrt{3}\sigma\mathcal{B}^{2}$, $\tilde{\nu}=i\sqrt{3}\nu\mathcal{B}^{-2}$,
$\tilde{\kappa}=i\sqrt{3}\kappa\mathcal{B}^{2}$, $\tilde{\epsilon}=\epsilon+\frac{1}{2}D\ln\mathcal{B}$,
$\tilde{\gamma}=\gamma+\frac{1}{2}\Delta\ln\mathcal{B}$, $\tilde{\beta}=\beta+\frac{1}{2}\delta\ln\mathcal{B}$, $\tilde{\alpha}=\alpha+\frac{1}{2}\delta^*\ln\mathcal{B}$. 

\section{An invariant formulation of the Bianchi identities}
\label{sec:newvect}

We define the following three vectors

\begin{subequations}
\label{eqn:tvectorsdef}
\begin{eqnarray}
T_{\mu}&=&n^{\nu}\nabla_{\mu}\ell_{\nu}+m^{\nu}\nabla_{\mu}\bar{m}_{\nu}+
\nabla_{\mu}\ln\mathcal{B}, \\
T^+_{\mu}&=&\mathcal{B}\cdot \ell^{\nu}\nabla_{\mu}m_{\nu}, \\
T^-_{\mu}&=&\mathcal{B}^{-1}\cdot n^{\nu}\nabla_{\mu}\bar{m}_{\nu};
\end{eqnarray}
\end{subequations}
which show the property of being invariant under a spin/boost transformation,
making them promising quantities to be related to curvature invariants. However, as we will
show later, this is not enough for such a goal. The reduced
spin coefficients can be written in terms of these vectors as

\begin{subequations}
\label{eqn:reduspinformula}
\begin{eqnarray}
\rho&=&-\mathcal{B}^{-1}\bar{m}^{\mu}T^+_{\mu}, \\
\tilde{\lambda}&=&i\sqrt{3}\mathcal{B}^{-1}\bar{m}^{\mu}T^-_{\mu}, \\
\tilde{\epsilon}&=&\frac{1}{2}\ell^{\mu}T_{\mu}, \\
\mu&=&\mathcal{B}m^{\mu}T^-_{\mu}, \\
\tilde{\sigma}&=&-i\sqrt{3}\mathcal{B}m^{\mu}T^+_{\mu}, \\
\tilde{\gamma}&=&\frac{1}{2}n^{\mu}T_{\mu}, \\
\tau&=&-\mathcal{B}^{-1}n^{\mu}T^+_{\mu}, \\
\tilde{\nu}&=&i\sqrt{3}\mathcal{B}^{-1}n^{\mu}T^-_{\mu}, \\
\tilde{\beta}&=&\frac{1}{2}m^{\mu}T_{\mu}, \\
\pi&=&\mathcal{B}\ell^{\mu}T^-_{\mu}, \\
\tilde{\kappa}&=&-i\sqrt{3}\mathcal{B}\ell^{\mu}T^+_{\mu}, \\
\tilde{\alpha}&=&\frac{1}{2}\bar{m}^{\mu}T_{\mu}.
\end{eqnarray}
\end{subequations}

Our aim is to find quantities that can be related to curvature invariants and/or 
functions of them. As such invariants do not depend on any of the tetrad transformations,
it is necessary to find quantities that show this same feature. The two transformations
that rotate the $\ell^{\mu}$ and $n^{\mu}$ have already been taken care of, by choosing
to be in a transverse frame. The spin/boost transformation has been taken care of 
in the definition of the $T$ vectors in Eq.~(\ref{eqn:tvectorsdef}) where it is evident that such vectors
are invariants under this third type of transformations. However, we still have to consider
a fourth type of tetrad transformation, namely the exchange operation $\ell^{\mu}\leftrightarrow n^{\mu}$ and
$m^{\mu} \leftrightarrow \bar{m}^{\mu}$. Such a transformation acts on these vectors
 in the following way

\begin{subequations}
\label{eqn:exchangeT}
\begin{eqnarray}
T_{\mu}&\leftrightarrow&-T_{\mu}, \\
T^+_{\mu}&\leftrightarrow&T^-_{\mu}.
\end{eqnarray}
\end{subequations}

The fact that these three vectors are not invariant under exchange transformation is 
obviously an obstacle to expressing them as functions of curvature invariants.
The problem is therefore defining
alternative quantities that are instead invariant under exchange transformations. In
order to do so, we introduce the following set of self-dual two-forms:

\begin{eqnarray}
\label{eqn:twoforms}
\Sigma_{\mu\nu}&=&2\ell_{[\mu}n_{\nu]}-2m_{[\mu}\bar{m}_{\nu]}, \\
\Sigma^{+}_{\mu\nu}&=&2\mathcal{B}\cdot\ell_{[\mu}m_{\nu]}, \\
\Sigma^-_{\mu\nu}&=&2\mathcal{B}^{-1}\cdot n_{[\mu}\bar{m}_{\nu]}. 
\end{eqnarray}

These two-forms are called self-dual because they are invariant under the operation
of hodge dual transformation. By definition, they are also invariant
under spin-boost transformations, while an exchange operation $\ell^{\mu}\leftrightarrow n^{\mu}$ and $m^{\mu}\leftrightarrow \bar{m}^{\mu}$ acts on them
in the following way

\begin{subequations}
\label{eqn:exchangeSigma}
\begin{eqnarray}
\Sigma_{\mu\nu}&\leftrightarrow&-\Sigma_{\mu\nu}, \\
\Sigma^{+}_{\mu\nu}&\leftrightarrow&\Sigma^{-}_{\mu\nu},
\end{eqnarray}
\end{subequations}
i.e., with similar transformation properties as those of the $T$ vectors.  
We can therefore construct quantities that
show invariance under exchange operations, as follows

\begin{subequations}
\label{eqn:modvectors}
\begin{eqnarray}
A_{\mu}&=&\Sigma^{+\nu}_{\mu}T^-_{\nu}+\Sigma^{-\nu}_{\mu}T^+_{\nu} ,\\
B_{\mu}&=&\Sigma^{+\nu}_{\mu}T^+_{\nu}+\Sigma^{-\nu}_{\mu}T^-_{\nu} ,\\
C_{\mu}&=&{\Sigma_{\mu}}^{\nu}T_{\nu}.
\end{eqnarray}
\end{subequations}

Given the properties of transformations under spin/boost and exchange operations of
the $\Sigma$ and $T$ variables, these
three vectors are invariant under both transformations. We expect these vectors
to be good candidates for relations involving only curvature invariants, as indeed we will
show to be the case.

The reduced spin coefficients can be easily expressed in terms of these newly 
introduced vectors

\begin{subequations}
\label{eqn:reduspinformulamodif}
\begin{eqnarray}
\rho=-\ell^{\mu}A_{\mu}, &  \mu=n^{\mu}A_{\mu}, \\
\tilde{\lambda}=i\sqrt{3}\ell^{\mu}B_{\mu}, & \tilde{\sigma}=-i\sqrt{3}n^{\mu}B_{\mu}, \\
\tilde{\epsilon}=-\frac{1}{2}\ell^{\mu}C_{\mu}, & \tilde{\gamma}=\frac{1}{2}n^{\mu}C_{\mu}, \\
\tau=-m^{\mu}A_{\mu}, & \pi=\bar{m}^{\mu}A_{\mu}, \\
\tilde{\nu}=i\sqrt{3}m^{\mu}B_{\mu}, & \tilde{\kappa}=-i\sqrt{3}\bar{m}^{\mu}B_{\mu}, \\
\tilde{\beta}=-\frac{1}{2}m^{\mu}C_{\mu}, &
\tilde{\alpha}=\frac{1}{2}\bar{m}^{\mu}C_{\mu}, 
\end{eqnarray}
\end{subequations}
i.e. they become simple contractions of the three 
vectors $A^{\mu}$,  $B^{\mu}$ and $C^{\mu}$ along the four null vectors. It is interesting
to see that in this case for example the four spin coefficients $\tilde{\lambda}$,
$\tilde{\sigma}$, $\tilde{\nu}$ and $\tilde{\kappa}$ are all directional derivatives of the
same vector $B^{\mu}$, and we will see that this is an important property to derive
the Goldberg-Sachs theorem in a simplified way.

We will now turn to the Bianchi identities, which, using the definitions of the three
vectors given in Eq.~(\ref{eqn:modvectors}) and Eq.~(\ref{eqn:bianchimod}), can be
expressed in a compact way as the following system of two equations

\begin{subequations}
\label{eqn:bianchiabc}
\begin{eqnarray}
A_{\mu}&=&\frac{1}{\Psi_-^2+3\Psi_+^2}\left(\sqrt{3}\Psi_+\mathcal{F}_{\mu}-i\Psi_-\mathcal{G}_{\mu}\right),  \label{eqn:bianchiabc1} \\
B_{\mu}&=&\frac{1}{\Psi_-^2+3\Psi_+^2}\left(\sqrt{3}\Psi_+\mathcal{G}_{\mu}-i\Psi_-\mathcal{F}_{\mu}\right), \label{eqn:bianchiabc2}
\end{eqnarray}
\end{subequations}
where

\begin{subequations}
\label{eqn:bianchireddeffg}
\begin{eqnarray}
\mathcal{F}_{\mu}&=&-\frac{\nabla_{\mu}\Psi_+}{\sqrt{3}}, \\
\mathcal{G}_{\mu}&=&2i\Psi_-C_{\mu}-i\nabla_{\mu}\Psi_-.
\end{eqnarray}
\end{subequations}

The eight Bianchi identities given in Eq.~(\ref{eqn:bianchimod}) can be derived by simply contracting these
two equations along the four null vectors constituting the Newman-Penrose tetrad.

Eq.~(\ref{eqn:bianchiabc}) is a very interesting way of rewriting the Bianchi identities. Some known results
follow very nicely from this approach to the NP formalism. For example, the Goldberg-Sachs
theorem can be seen as a straightforward consequence of Eq.~(\ref{eqn:bianchiabc2}). Indeed,
in the case of Petrov type D limit, one has that $\Psi_-\rightarrow 0$, and using Eq.~(\ref{eqn:bianchiabc2})
this implies simply that $B^{\mu}\rightarrow0$, and therefore the four spin coefficients
$\tilde{\lambda}$,
$\tilde{\sigma}$, $\tilde{\nu}$ and $\tilde{\kappa}$ vanish in this limit, which is what the
theorem states. 

The identity in Eq.~(\ref{eqn:bianchiabc1}) can also be studied in the Petrov type D limit, 
giving that the field $A^{\mu}$ tends to the value

\begin{equation}
\label{eqn:limitA}
A_{\mu} \rightarrow -\frac{1}{3}\nabla_{\mu}\ln\Psi_+.
\end{equation}
Such a limit is consistent with the well known
expressions of the
four spin coefficients $\rho$, $\mu$, $\tau$ and $\pi$ in Kerr space-time, which
are given in Boyer-Lindquist coordinates by

\begin{subequations}
\label{eqn:limitrho}
\begin{eqnarray}
\rho = -\frac{1}{r-ia\cos\theta}, & \mu = \rho^2\rho^*\Gamma/2, \\
\tau = -ia\rho\rho^*\sin\theta/\sqrt{2}, & \pi = ia\rho^2\sin\theta/\sqrt{2},
\end{eqnarray}
\end{subequations}
where $\Gamma=r^2-2Mr+a^2$ and $M$ and $a$ are the mass and angular momentum 
per unit mass of the black hole, respectively.

Eq.~(\ref{eqn:bianchiabc}) suggests that the Bianchi identities can be used as eliminant 
conditions to determine, once the curvature invariants are given, and therefore $\Psi_+$
and $\Psi_-$ are given, the fields $A_{\mu}$ and $B_{\mu}$, provided the field $C_{\mu}$
is given too. However, one of the three fields seems to be undetermined, contrasting
with the fact that we have fixed all the gauge degrees of freedom. We would in fact 
expect all these
quantities to be functions of the curvature invariants,
or derivatives of them, or of some other invariant contractions of the Weyl tensor, given
that we have fixed the tetrad completely. Nonetheless this
apparent lack of information can be explained with the fact that we still
have a limited vision of the whole formalism, since the Ricci identities have not been
analyzed in a way that is consistent with this new approach. We expect these identities
to give the missing information and the result will be presented in a follow-up paper.


\section{Conclusions}
\label{sec:concl}

We have shown that in transverse tetrads it is possible to write the curvature
degrees of freedom as simple functions of curvature invariants and, moreover,
by introducing three new vector fields, it is possible to rewrite the Bianchi identities
in the Newman-Penrose formalism as a simple set of two equations, relating
these three vector fields to the curvature invariants. In fact, in this new picture the Bianchi identities can
be thought as simple eliminant relations that give two of the newly
introduced vector fields as functions of curvature invariants and of a third, seemingly
undetermined, field. We expect to complete the information by analyzing the Ricci
identities within this new approach, which will be the subject of the next paper on the
topic.

We expect this approach to be very promising for a better understanding of the NP 
formalism, reducing considerably the complexity of the equations, also in view of a
possible extension of these concepts to higher dimensions \cite{Coley:2004jv, Pravda:2004ka, Pravdova:2005ey, Durkee:2010xq, Godazgar:2010ks}, where a lot of work is being
done in order to generalize the Teukolsky perturbative approach \cite{Durkee:2010qu, Durkee:2010ea}, and
numerical simulations \cite{Zilhao:2010sr, Witek:2010xi, Witek:2010az, Zilhao:2011yc, Nakao:2009dc, Yoshino:2009xp} are starting to explore such dynamical scenarios.


\acknowledgments

It is a pleasure to thank Vitor Cardoso, Leonardo Gualtieri and Ulrich Sperhake for
useful discussions and for careful proofreading of this manuscript.
The author has been funded by the Funda\c c\~ao para a Ci\^encia e Tecnologia 
through grant SFRH/BPD/47955/2008 and through projects CERN/FP/109290/2009,
PTDC/FIS/098025/2008 and PTDC/FIS/098032/2008. This work was supported by
the DyBH0-256667 ERC Starting Grant.


\bibliographystyle{apsrev}
\bibliography{references}


\end{document}